\documentclass[epj]{svjour}
\usepackage{graphicx}  
\usepackage{bm}  
\newcommand{\beq}{\begin{equation}}
\newcommand{\eeq}{\end{equation}}
\newcommand{\bea}{\begin{eqnarray}}
\newcommand{\eea}{\end{eqnarray}}       

\newcommand{\simlt}{\stackrel{<}{{}_\sim}}
\newcommand{\simgt}{\stackrel{>}{{}_\sim}}

\newcommand{\bit}{\begin{itemize}}
\newcommand{\eit}{\end{itemize}}

\begin{document}
\title{Nucleon Form Factors in Dispersion Theory}
\author{H.-W. Hammer
}                     
%
%
\institute{Helmholtz-Institut f{\"u}r
Strahlen- und Kernhysik (Theorie), Universit\"at Bonn, 
Nussallee 14-16, D-53115 Bonn, Germany}
\date{Received: date / Revised version: date}
%
\abstract{
Dispersion relations provide a powerful tool to analyse the
electromagnetic form factors of the nucleon both in the space-like
and time-like regions with constraints from other experiments,
unitarity, and perturbative QCD. We give a brief introduction into 
dispersion theory for nucleon form factors and present first results
from our ongoing form factor analysis. We also calculate the 
two-pion continuum contribution to the isovector spectral functions
drawing upon the new high statistics measurements of the pion form
factor by the CMD-2, KLOE, and SND collaborations.
} 
\titlerunning{Nucleon Form Factors in Dispersion Theory}
\authorrunning{H.-W. Hammer}
\maketitle

\section{Introduction}
\label{sec:intro}

The electromagnetic form factors of the nucleon offer a unique window
on strong interaction dynamics over a wide range of momentum
transfers \cite{Gao:2003ag,Hyde-Wright:2004gh}.
At small momentum transfers, they are sensitive
to the gross properties of the nucleon like the charge and magnetic moment,
while at high momentum transfers they encode information on
the quark substructure of the nucleon as described by QCD.

Their detailed understanding is important for unraveling aspects of 
perturbative and nonperturbative nucleon structure. 
The form factors also contain
important information on nucleon radii and vector meson coupling
constants. Moreover, they are an
important ingredient in a wide range of experiments from
Lamb shift measurements \cite{Ude97} to measurements of the
strangeness content of the nucleon \cite{strange}.

With the advent of the new continuous beam electron accelerators
such as CEBAF (Jefferson Lab.), ELSA (Bonn), and MAMI (Mainz), a
wealth of precise data for space-like momentum transfers
has become available \cite{Ostrick}. 
Due to the difficulty of the experiments, the
time-like form factors are less well known. While there is a fair 
amount of information on the proton time-like form factors 
\cite{protont1,protont2,protont3,protont4,protont5},
only one measurement of the neutron form factor from the pioneering 
FENICE experiment \cite{Ant98} exists.

It has been known for a long time that the pion plays an important role in 
the long-range structure of the nucleon \cite{FHK}. This connection
was made more precise using dispersion theory in the 1950's
\cite{CKGZ58,FGT58}. Subsequently, Frazer and Fulco have written down 
partial wave dispersion relations that relate the nucleon electromagnetic 
structure to pion-nucleon ($\pi N$) scattering and predicted the existence of
the $\rho$ resonance \cite{FF,FF60a}.
Despite of this success, the central role of the $2\pi$ continuum
in the isovector spectral function has often been ignored.
H\"ohler and Pietarinen pointed out that this omission leads to
a gross underestimate of the isovector radii of the nucleon \cite{HP2}.
They first performed a consistent 
dispersion analysis of the electromagnetic form factors of the
nucleon \cite{Hohler:1976ax} including the $2\pi$ continuum
derived from the pion form factor and $\pi N$-scattering data \cite{HP}.
In the mid-nineties, this analysis has been updated 
by Mergell, Mei\ss ner, and Drechsel \cite{MMD96} and was later
extended to include data in the time-like region \cite{HMD96,slacproc}. 
Recently, the new precise data for the neutron electric form factor have 
been included as well \cite{HM04}.

Using chiral perturbation theory (ChPT), the long-range pionic structure
of the nucleon can be connected to the Goldstone boson dynamics of QCD
\cite{BKMrev}. The nonresonant part of
the $2\pi$ continuum is in excellent agreement with the
phenomenological analysis \cite{BKMspec} and the $\rho$-meson
contribution can be included as well \cite{KM,Norbert,Schindler:2005ke}.
It is well known that vector mesons play an important 
role in the electromagnetic structure of the nucleon, see e.g. 
Refs.~\cite{FF,Sak,GoSa,Gari,Lomo,Dubni}, and the remaining contributions
to the spectral function
have usually been approximated by vector meson resonances. 

A new twist to this picture was recently given by Fried\-rich and Walcher 
\cite{FW}. They interpreted the form factor data based 
on a phenomenological fit with an ansatz for the pion cloud using the 
idea that the proton can be thought of as virtual neutron-positively 
charged pion pair. A very long-range contribution to the charge distribution 
in the Breit frame extending out to about 2~fm was found and attributed to 
the pion cloud. This was shown to be in conflict with the phenomenologically
known $2\pi$ continuum and ChPT by Hammer, Drechsel, and Mei\ss ner
\cite{Hammer:2003qv}. We will address this conundrum
in more detail in Sec.~\ref{sec:picloud}.

In this talk we give a brief introduction into dispersion theory for 
nucleon form factors and present preliminary results
from our ongoing form factor analysis. We also calculate the 
two-pion continuum contribution to the isovector spectral functions
drawing upon the new high statistics measurements of the pion form
factor by the CMD-2, KLOE, and SND collaborations.
Finally we address the question of the range of the pion cloud and
give an outlook on future work.

\section{Definitions}
\label{sec:def}

\begin{figure}[ht] 
\centerline{\includegraphics*[width=4.5cm,angle=0]{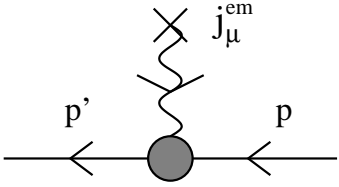}}
\caption{
The nucleon matrix element of the electromagnetic current $j_\mu^{\rm em}$.}
\label{fig:curr}
\end{figure}

The electromagnetic (em) structure of the nucleon is determined by
the matrix element of the current operator $j_\mu^{\rm em}$ 
between nucleon states as illustrated in Fig.~\ref{fig:curr}.

Using Lorentz and gauge invariance, this matrix element can  
be expressed in terms of two form factors,
\begin{equation}
\langle p' | j_\mu^{\rm em} | p \rangle = \bar{u}(p')
\left[ F_1 (t) \gamma_\mu +i\frac{F_2 (t)}{2 M} \sigma_{\mu\nu}
q^\nu \right] u(p)\,,
\end{equation}
where $M$ is the nucleon mass and $t=(p'-p)^2$ 
the four-momentum transfer. For data in the space-like region, it is
often convenient to use the variable $Q^2=-t>0$.
The functions
$F_1(t)$ and $F_2(t)$ are the Dirac and Pauli form factors, respectively.
They are normalized at $t=0$ as
\begin{equation}
\label{norm}
F_1^p(0) = 1\,, \; F_1^n(0) = 0\,, \; F_2^p(0) =  \kappa_p\,,
\; F_2^n(0) = \kappa_n\, ,
\end{equation}
with $\kappa_p=1.79$ and $\kappa_n=-1.91$ the anomalous magnetic moments of
protons and neutrons in nuclear magnetons, respectively.

It is convenient to work in the isospin basis and to 
decompose the form factors into isoscalar and isovector parts,
\begin{equation}
F_i^s = \frac{1}{2} (F_i^p + F_i^n) \, , \quad
F_i^v = \frac{1}{2} (F_i^p - F_i^n) \, ,
\end{equation}
where $i = 1,2 \,$. 

The experimental data are usually given for the Sachs form factors
\begin{eqnarray}
\label{sachs}
G_{E}(t) &=& F_1(t) - \tau F_2(t) \, , \\
G_{M}(t) &=& F_1(t) + F_2(t) \, , \nonumber
\end{eqnarray}
where $\tau = -t/(4 M^2)$.
In the Breit frame, $G_{E}$ and $G_{M}$ may be interpreted as
the Fourier transforms of the charge and magnetization distributions,
respectively.              

The nucleon radii can be defined from the low-$t$ expansion
of the form factors,
\beq
F(t)=F(0)\left[1+t\langle r^2 \rangle /6 +\ldots \right]\,,
\eeq
where $F(t)$ is a generic form factor. In the case of the electric
and Dirac form factors of the neutron, $G_E^n$ and $F_1^n$, the 
normalization factor $F(0)$ is simply dropped.

\section{Dispersion Relations and Spectral Decomposition}
\label{sec:specdeco}

Based on unitarity and analyticity, dispersion relations relate
the real and imaginary parts of the electromagnetic (em) nucleon form factors. 
Let $F(t)$ be a generic symbol for any one of the four independent 
nucleon form factors. We write down an unsubtracted
dispersion relation of the form
\begin{equation}
\label{disprel}
F(t) = \frac{1}{\pi} \, \int_{t_0}^\infty \frac{{\rm Im}\, 
F(t')}{t'-t-i\epsilon}\, dt'\, ,
\label{emff:disp} 
\end{equation}
where $t_0$ is the threshold of the lowest cut of $F(t)$ (see below)
and the $i\epsilon$ defines the integral for values of $t$ on the 
cut.\footnote{The convergence of an unsubtracted dispersion relation
for the form factors has been assumed. We could have used
a once subtracted dispersion relation as well
since the normalization of the form factors is known.}
Eq.~(\ref{emff:disp}) relates the em structure
of the nucleon to its absorptive behavior.

The imaginary part ${\rm Im}\, F$ entering Eq.~(\ref{disprel}) 
can be obtained from a spectral decomposition \cite{CKGZ58,FGT58}. 
For this purpose it is most convenient to consider the 
em current matrix element in the time-like region ($t>0$), which is 
related to the space-like region ($t<0$) via crossing symmetry.
The matrix element can be expressed as
\begin{eqnarray}
\label{eqJ}
J_\mu &=& \langle N(p) \overline{N}(\bar{p}) | j_\mu^{\rm em}(0) | 0 \rangle \\
&=& \bar{u}(p) \left[ F_1 (t) \gamma_\mu +i\frac{F_2 (t)}{2 M} \sigma_{\mu\nu}
(p+\bar{p})^\nu \right] v(\bar{p})\,,\nonumber
\end{eqnarray}
where $p$ and $\bar{p}$ are the momenta of the nucleon and an\-ti\-nuc\-le\-on
created by the current $j_\mu^{\rm em}$, respectively. 
The four-momentum transfer in the time-like region is $t=(p+\bar{p})^2$. 

\begin{figure}[ht] 
\centerline{\includegraphics*[width=7.5cm,angle=0]{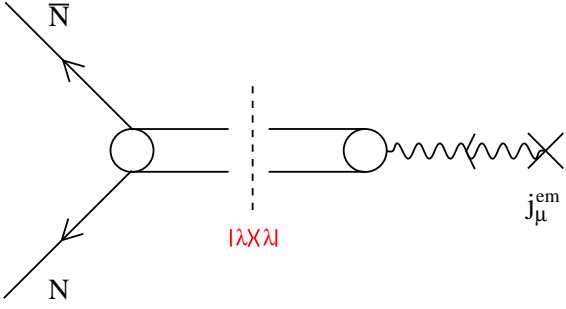}}
\caption{
The spectral decomposition of the 
nucleon matrix element of the electromagnetic current $j_\mu^{\rm em}$.}
\label{fig:spec}
\end{figure}

Using the LSZ reduction formalism, the imaginary part
of the form factors is obtained by inserting a complete set of
intermediate states as \cite{CKGZ58,FGT58}
\begin{eqnarray}
\label{spectro}
{\rm Im}\,J_\mu &=& \frac{\pi}{Z}(2\pi)^{3/2}{\cal N}\,\sum_\lambda
 \langle p | \bar{J}_N (0) | \lambda \rangle \qquad\\
& &\times\langle \lambda | j_\mu^{\rm em} (0) | 0 \rangle \,v(\bar{p})
\,\delta^4(p+\bar{p}-p_\lambda)\,,\nonumber
\end{eqnarray}
where ${\cal N}$ is a nucleon spinor normalization factor, $Z$ is
the nucleon wave function renormalization, and $\bar{J}_N (x) =
J^\dagger(x) \gamma_0$ with $J_N(x)$ a nucleon source.
This decomposition is illustrated in Fig.~\ref{fig:spec}.
It relates the spectral function to on-shell matrix elements of other
processes.

The states $|\lambda\rangle$ are asymptotic states of
momentum $p_\lambda$ which are stable with respect to the strong 
interaction. They must carry the same quantum numbers as as
the current $j^{\rm em}_\mu$: $I^G(J^{PC})=0^-(1^{--})$ for
the isoscalar current and $I^G(J^{PC})=1^+(1^{--})$ for the
isovector component of $j^{\rm em}_\mu$. 
Furthermore, they have no net baryon number. Because of $G$-parity, states
with an odd number of pions only contribute to the iso\-scalar
part, while states with an even number contribute to the 
isovector part.
For the isoscalar part  the lowest mass states are: $3\pi$,
$5\pi$, $\ldots$, $K\bar{K}$, $K\bar{K}\pi$, $\ldots$; 
for the isovector part they are: $2\pi$, $4\pi$, $\ldots$. 

Associated with each intermediate state is a
cut starting at the corresponding threshold in $t$ and running to
infinity. As a consequence,
the spectral function ${\rm Im}\, F(t)$ is different from zero along the
cut from $t_0$ to $\infty$ with $t_0 = 4 \, (9) \, M_\pi^2$ for the
isovector (isoscalar) case.

The spectral functions are the central quantities in the 
dispersion-theoretical approach. Using Eqs.~(\ref{eqJ},\ref{spectro}), they
can in principle be constructed from experimental data. 
In practice, this program can only be carried out for 
the lightest two-particle intermediate states ($2\pi$ and $K\bar{K}$)
\cite{HP,Hammer:1998rz,Hammer:1999uf}.

The longest-range (and therefore most important at low momentum
transfer) pion cloud contribution comes from the $2\pi$ intermediate state 
in the  isovector form factors. A new calculation of this contribution 
will be discussed in the following section.

\section{Two--Pion Continuum}
\label{sec:2pico}

In this section, we re-evaluate the $2\pi$ contribution in a 
model--independent way \cite{Belushkin:2005ds}
using the latest experimental data for the pion
form factor from CMD-2 \cite{CMD2}, KLOE \cite{KLOE}, and
SND \cite{SND}.

We follow  Ref.~\cite{LB} and express the $2\pi$ contribution to the
the isovector spectral functions  in terms of the pion 
charge form factor $F_\pi (t)$ and the P--wave $\pi\pi \to \bar N N$ 
amplitudes $f^1_\pm(t)$. The $2\pi$ continuum is expected to be the
dominant contribution to the isovector spectral function 
from threshold up to masses of about 1~GeV \cite{LB}.
Here, we use the expressions
\bea 
\nonumber 
{\rm Im}~G_E^{v} (t) &=& \frac{q_t^3}{M\sqrt{t}}\, 
F_\pi (t)^* \, f^1_+ (t)~,\\ 
{\rm Im}~G_M^{v} (t) &=& \frac{q_t^3}{\sqrt{2t}}\, 
F_\pi (t)^* \, f^1_- (t)~, 
\label{uni} 
\eea 
where $q_t=\sqrt{t/4-M_\pi^2}$. The imaginary parts of the Dirac
and Pauli Form factors can be obtained using Eq.~(\ref{sachs}).

The  P--wave $\pi\pi \to \bar N N$ amplitudes $f_\pm^1(t)$ are tabulated in  
Ref.~\cite{LB}. (See also Ref.~\cite{Pieta} for an unpublished
update that is consistent with Ref.~\cite{LB}.) 
We stress that the representation of Eq.~(\ref{uni}) 
gives the exact isovector spectral functions for $4M_\pi^2 \leq t 
\leq 16 M_\pi^2$, but in practice holds up to $t \simeq 50 M_\pi^2$.
Since the contributions from $4\pi$ and higher 
intermediate states is small up to  $t \simeq 50 M_\pi^2$, $F_\pi(t)$
and the $f_\pm^1(t)$ share the same phase in this region and the 
two quantities can be replaced by their absolute values.\footnote{
We note that representation of Eq.~(\ref{uni}) is most useful for
our purpose. The manifestly real functions $J_\pm (t) = f_\pm^1(t)/ 
F_\pi (t)$ also tabulated in Ref.~\cite{LB}
contain assumptions about the pion form factor
which leads to inconsistencies when used together with the 
updated $F_\pi (t)$.}

\begin{figure}[t] 
\centerline{\includegraphics*[width=8.5cm,angle=0]{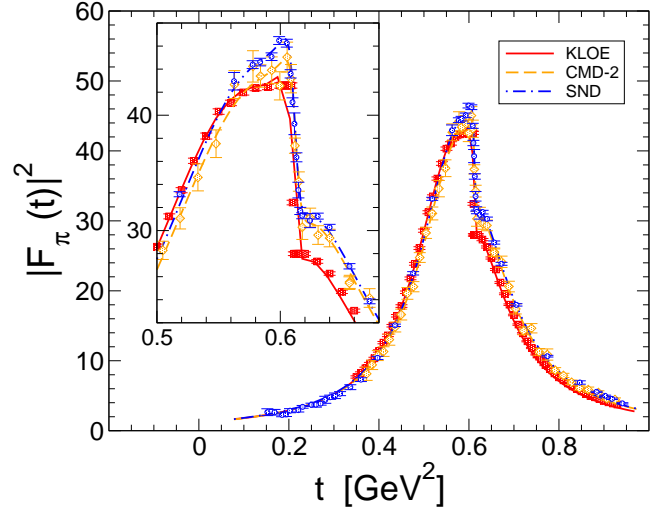}}
\caption{
The pion electromagnetic form factor $F_\pi (t)$ in the time-like
region as a function of the momentum transfer $t$. The diamonds,
squares, and circles show the high statistics data from the 
CMD-2 \cite{CMD2}, KLOE \cite{KLOE}, and SND \cite{SND}
collaborations, respectively. The dashed, solid, and dash-dotted 
lines are our model parametrizations. The inset shows
the discrepancy in the resonance region in more detail.
}
\label{piFF}
\end{figure}

The updated pion form factor is shown in Fig.~\ref{piFF}.
The diamonds, squares, and circles show the high statistics data from the 
CMD-2 \cite{CMD2}, KLOE \cite{KLOE}, and SND \cite{SND}
collaborations, respectively. The dashed, solid, and dash-dotted 
lines are our model parametrizations which are of the Gounaris-Sakurai type
\cite{MMD96,GoSa}. The form factor shows a pronounced 
$\rho$-$\omega$ mixing in the vicinity of the $\rho$-peak.
There are discrepancies between the three experimental data sets for the 
pion form factor \cite{SND}. The discrepancies in the 
$\rho$-resonance region are  shown in more detail in the inset
of Fig.~\ref{piFF}. Since we are not in the 
position to settle this experimental problem, we will take the
three data sets at face value. We will evaluate the  $2\pi$ continuum 
given by Eq.~(\ref{uni})
for all three sets and estimate the errors from the discrepancy 
between the sets.

\begin{figure}[ht] 
\centerline{\includegraphics*[width=8.5cm,angle=0]{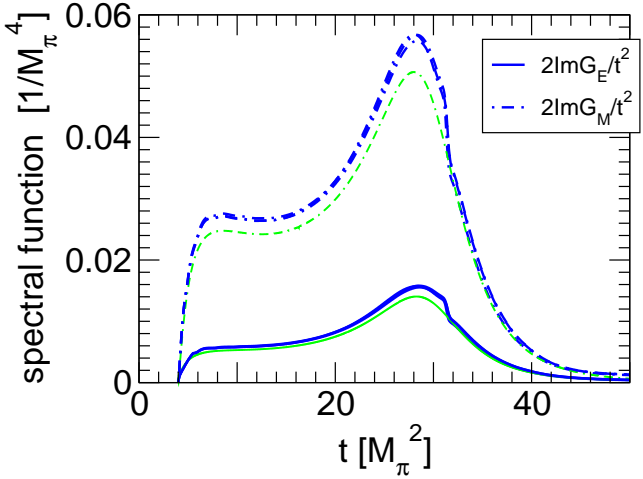}}
\caption{
The $2\pi$ spectral function using the new high statistics data for 
the pion form factor \cite{CMD2,KLOE,SND}. The spectral functions weighted by 
$1/t^2$ are shown for $G_E$ (solid line) and $G_M$ (dash-dotted line)
in units of $1/M_\pi^4$. 
The previous results by H\"ohler et al.~\cite{LB} (without $\rho$-$\omega$ 
mixing) are shown for comparison by the gray/green lines.}
\label{2pispec}
\end{figure} 
Using the new high statistics pion form factor data \cite{CMD2,KLOE,SND}
and the amplitudes $f_\pm^1(t)$
tabulated in  Ref.~\cite{LB}, we obtain the spectral functions
shown in Fig.~\ref{2pispec}  \cite{Belushkin:2005ds}.
We show the spectral functions weighted by $1/t^2$ for 
$G_E$ (solid line) and $G_M$ (dash-dotted line). The previous 
results by H\"ohler et al.~\cite{LB} (without $\rho$-$\omega$ 
mixing) are given for comparison  by the gray/green lines.
The general structure of the two
evaluations is the same, but there is a difference in magnitude of about 10\%.
The difference between the three data sets for the pion form factor is
very small and indicated by the line thickness. The difference in the
form factors is largest in the $\rho$-peak region (cf.~Fig.~\ref{piFF}),
but this region is suppressed by the $\pi\pi \to \bar N N$ amplitudes 
$f_\pm^1(t)$ which show a strong fall-off as $t$ increases.

The spectral functions have two distinct features. First, as already 
pointed out in \cite{FF}, they contain the important contribution of 
the $\rho$-meson with its peak at $t \simeq 30 M_\pi^2$. 
Second, on the left shoulder of the $\rho$, the isovector spectral functions 
display a very pronounced enhancement close to the two-pion threshold. This 
is due to the logarithmic singularity on the second Riemann sheet located at 
$t_c = 4M_\pi^2 - M_\pi^4/M^2 = 3.98 M_\pi^2$, very close to the threshold. 
This pole comes from the projection of the nucleon Born graphs, or in modern 
language, from the triangle diagram. 

If one were to neglect this important unitarity correction, one would 
severely  underestimate the nucleon isovector radii \cite{HP2},
\begin{equation}
\langle r^2\rangle_i^{v} = \frac{6}{\pi}
 \int_{4M_\pi^2}^\infty \frac{dt}{{t}^2} \, {\rm Im}\,
  G_i^{v} (t) \, ,
\end{equation}
where $i=E,M$. In fact, precisely the 
same effect is obtained at leading one-loop accuracy  in 
relativistic chiral perturbation 
theory \cite{GSS,UGMlec}. This topic was also discussed in 
heavy baryon ChPT \cite{BKMspec,Norbert} and in a covariant 
calculation based on infrared regularization \cite{KM}. Thus, the most 
important $2\pi$ contribution to the nucleon form factors can be determined 
by using either unitarity or ChPT (in the latter case, of course, the $\rho$ 
contribution is not included).

\section{Spectral Functions}
\label{sec:specf}

As discussed above the spectral function can at present only be obtained
from unitarity arguments for 
the lightest two-particle intermediate states ($2\pi$ and $K\bar{K}$)
\cite{HP,Hammer:1998rz,Hammer:1999uf}.
The $\rho\pi$ continuum contribution can be obtained from the Bonn-J\"ulich
model \cite{Meissner:1997qt}.

The remaining contributions can be parametrized by vector meson
poles. On one hand, 
the lower mass poles can be identified with physical vector 
mesons such as the $\omega$ and the $\phi$. In the the case of the 
$3\pi$ continuum, e.g., it has been shown in ChPT that the nonresonant
contribution is very small and the spectral function is dominated 
by the $\omega$ \cite{BKMspec}.
The higher mass poles on the other hand, are simply an effective way
to parametrize higher mass strength in the spectral function.

\begin{figure}[ht] 
\centerline{\includegraphics*[width=8cm,angle=0]{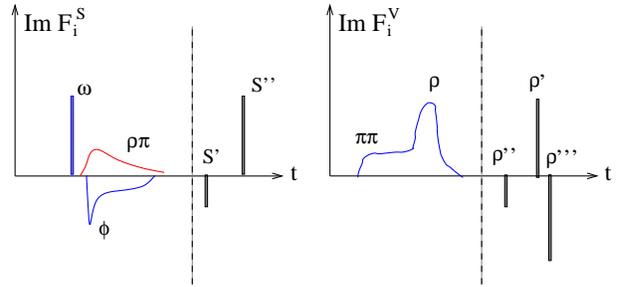}}
\caption{Illustration of the spectral function used in the
dispersion analysis. The vertical dashed line separates the 
well-known low-mass contributions ($2\pi$, $K\bar{K}$, and  
$\rho\pi$ continua as well as the $\omega$ pole)
from the effective poles at higher momentum transfers.
}
\label{specillu}
\end{figure} 

For our current best fit, the spectral function includes the 
$2\pi$, $K\bar{K}$, and  $\rho\pi$ continua from unitarity
and the  $\omega$ pole. In addition to that there are a number
of effective poles at higher momentum transfers
in both the isoscalar and isovector channels.
The spectral function then has the general structure
\bea
{\rm Im }\,F_i^{s} (t) &=& {\rm Im }\,F_i^{K\bar{K}} (t)
+ {\rm Im }\,F_i^{\rho\pi} (t) \nonumber \\
&& + \sum_{V=\omega,s_1,...} \pi a_i^{V}
\delta (M^2_{V}-t) \,, \quad i = 1,2 \, ,
\label{emff:s} \\
{\rm Im }\,F_i^{v} (t) &=& {\rm Im }\,F_i^{2\pi} (t)\nonumber \\
&& + \sum_{V= v_1,...} \pi a_i^{V}
\delta (M^2_{V}-t) \,, \quad i = 1,2 \, .
\label{emff:v}
\eea
which is illustrated in in Fig.~\ref{specillu}.
The vertical dashed line separates the well-known
low-mass contributions to the spectral function 
from the effective poles at higher momentum transfers.

In our fits, we also include the widths of the vector mesons.
The width and mass of the $\omega$ are taken from the particle
data tables while the masses and widths of the effective poles are
fitted to the form factor data. We have performed various fits with 
different numbers of effective poles and including/excluding some of the 
continuum contributions. In Sec.~\ref{sec:fits}, we will discuss
preliminary results of this ongoing effort.

\section{Constraints}
\label{sec:con}

The number of parameters in the fit function is reduced by enforcing
various constraints.
The first set of constraints concerns the low-$t$ behavior of the
form factors.
First, we enforce the correct normalization of the form factors,
which is given in Eq.~(\ref{norm}). Second, we constrain the 
neutron radius from a low-energy neutron-atom scattering experiment
\cite{Kop95,Kop97}.

Perturbative QCD (pQCD) constrains the behavior of the nucleon
em form factors for large momentum transfer.
Brodsky and Lepage \cite{BrL80} find for $t \to -\infty$,
\begin{equation}
F_i (t) \to (-t)^{-(i+1)} \, \left[ \ln\left(\frac{-t}{Q_0^2}\right)
\right]^{-\gamma} \, , \quad i = 1,2 \, ,
\label{emff:fasy1}
\end{equation}
where $Q_0 \simeq \Lambda_{\rm QCD}$.
The anomalous dimension $\gamma$ depends weakly on the number of
flavors, $\gamma = 2.148$, $2.160$, $ 2.173$ for $N_f = 3$, $4$,
$5$, in order.

The power behavior of the form factors at large $t$ can be easily 
understood from perturbative gluon exchange. In order to distribute the 
momentum transfer from the virtual photon
to all three quarks in the nucleon, at least two massless
gluons have to be exchanged. Since each of the gluons has a propagator 
$\sim 1/t$, the form factor has to fall off as $1/t^2$. In the case
of $F_2$, there is additional suppression by $1/t$ since a quark spin 
has to be flipped.
The power behavior of the form factors
leads to superconvergence relations of the form
\begin{equation}
\int_{t_0}^\infty {\rm Im}\, F_i (t) \;t^n dt =0\,,
\end{equation}
with $n=0$ for $F_{1}$ and $n=0,1$ for $F_{2}$.
The asymptotic behavior of Eq.~(\ref{emff:fasy1}) is obtained by
choosing the residues of the vector meson pole terms such that the
leading terms in the $1/t$-expansion cancel. 

The logarithmic term in Eq.~(\ref{emff:fasy1}) was included in
some of our earlier analyses \cite{MMD96,HMD96,HM04} but has
little impact on the fit. The particular way this constraint was
implemented, however, lead to an unphysical logarithmic singularity 
of the form factors in the time-like region.
In order to be able to include the data for the form factors at 
large time-like momentum transfers, the logarithmic constraint is not 
enforced in the current analysis.

The number of effective poles in
Eqs.~(\ref{emff:s}, \ref{emff:v}) is determined
by the stability criterion discussed in detail in \cite{Sab80}.
In short, we take the minimum number of poles necessary to fit the data.
For the preliminary results discussed in the next section, we took
4 effective isoscalar poles and 3 effective isovector poles whose
residua, masses, and widths are fitted to the data.
The number of free parameters is strongly reduced by the 
various constraints (unitarity, normalizations, superconvergence
relations), so that we end up with 19 free parameters in the 
preliminary fit presented in the next section. Our general strategy
is to reduce the number of parameters even further without sacrificing
the quality of the fit.

\section{Fit Results}
\label{sec:fits}

We now discuss some preliminary fit results that are representative for
the current status of the analysis. We present results for a fit
with 4 effective isoscalar poles and 3 effective isovector poles whose
residua, masses, and widths are fitted to the data.

\begin{figure*}[ht] 
\centerline{\includegraphics*[width=15cm,angle=0]{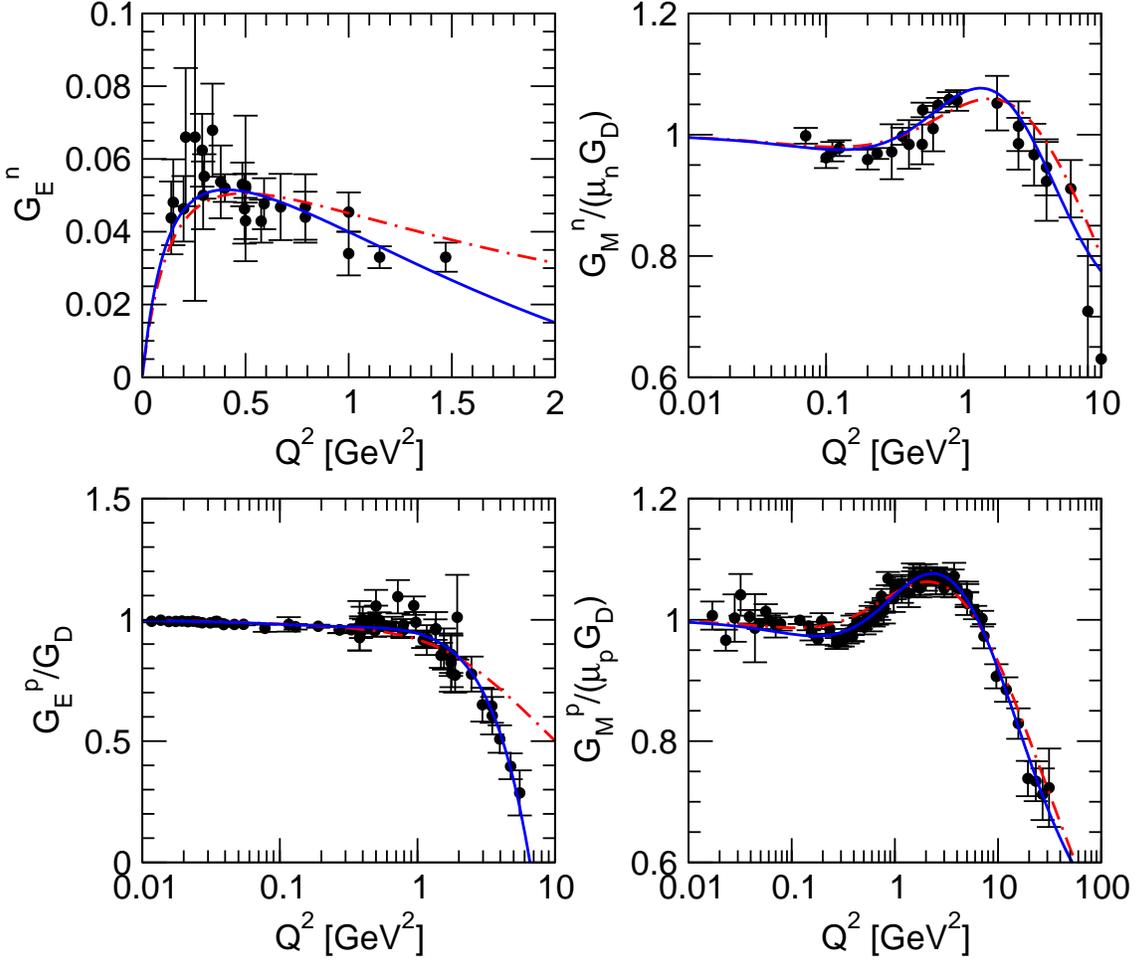}}
\caption{The nucleon electromagnetic form factors for space-like
momentum transfer.
The results for $G_M^n$, $G_E^p$, $G_M^p$ are normalized to the dipole fit.
The dash-dotted line gives the result of Ref.~\cite{HM04}, while
the the solid line indicates our {\it preliminary best fit}.
}
\label{fit2.1.2}
\end{figure*} 

In Fig.~\ref{fit2.1.2}, we show the results  for all four form factors 
compared to the world data for the form factors.
Our data basis is taken from Ref.~\cite{FW} and in addition 
also includes the new data that have appeared since 2003
(see Ref.~\cite{Ostrick}).
The results for $G_M^n$, $G_E^p$, $G_M^p$ are normalized to the 
phenomenological dipole fit:
\beq
\label{dipole}
G_D( Q^2)=\left(1+\frac{Q^2}{m_D^2}\right)^{-2} \,,
\eeq
where $m_D^2=0.71$ GeV$^2$.
The dash-dotted line gives the result of Ref.~\cite{HM04}, while
the the solid line indicates our present best fit.
The new fit leads to an improved description of the form factor data
compared with Ref.~\cite{HM04}. In particular, the rapid fall-off of
the JLab polarization data for $G_E^p$  \cite{Jones,Gayou}
is now described. The  $\chi^2$ per degree of freedom is
0.84. Note that we do not obtain a pronounced bump structure in
$G_E^n$ as observed in  Ref.~\cite{FW}. We will come back to this question
in Sec.~\ref{sec:picloud} and discuss the modifications in the spectral
function required to produce this structure.

The stability constraint requires to use the minimum number of 
poles required to describe the data \cite{Sab80}.
In the future, we plan to further reduce the number of effective
poles in order to improve the stability.

\begin{table}[ht]
\begin{tabular}{|c|c|c|c|}
\hline 
& this work & Ref.~\cite{HM04} & recent determ.\\
\hline\hline
$r_E^p$ [fm] & 0.84...0.857 & 0.848 & 0.886(15) 
\cite{Rosenfelder:1999cd,Sick:2003gm,Melnikov:1999xp}\\
$r_M^p$ [fm] & 0.85...0.875 & 0.857 & 0.855(35) 
\cite{Sick:2003gm,Sick:private} \\
$r_E^n$ [fm] & -0.12...-0.10 & -0.12 & -0.115(4) \cite{Kop97} \\
$r_M^n$ [fm] & 0.86...0.88 & 0.879 & 0.873(11) \cite{Kubon:2001rj}\\
\hline
\end{tabular}
\caption{\label{tab:nucrad}
Nucleon radii in fm extracted from the 
fit in Fig.~\ref{fit2.1.2}.}
\end{table}
In Table \ref{tab:nucrad}, we give the nucleon radii extracted from
our  fit. The neutron radius is included as a soft constraint in our
fit and therefore not a 
prediction.\footnote{A soft constraint is not implemented
exactly but deviations from the constraint are penalized in the $\chi^2$
of the fit.}
The other nucleon radii are generally in good agreement with
other recent determinations using only low-momentum-transfer data 
given in the table. Our result for the proton radius, however,
is somewhat small. This was already the case in the dispersion
analyses of Refs.~\cite{MMD96,HM04}. 
We speculate that the reason for this discrepancy
lies in inconsistencies in the data sets. In this type of global analysis
all four form factors are analyzed simultaneously and both data
at small and large momentum transfers enter. This can be an advantage
and disadvantage depending on the question at hand.
Another possible reason for the discrepancy is $2\gamma$ physics which was
neglected in the data analysis of most older experiments 
\cite{Guichon:2003qm}.

\section{Time-Like Data}
\label{sec:time}

We have also performed first fits that include data in the time-like
region. The extraction of these data is more challenging
than in the space-like region.
At the nucleon-antinucleon threshold, the electric and magnetic
form factors are equal by definition:
$G_{M} (4 M^2) = G_{E} (4 M^2)$,
while one expects the magnetic form factor to
dominate at large momentum transfer. Moreover, the form factors are complex
in the time-like region, since several physical thresholds are open.
Separating $|G_{M}|$ and $| G_{E}|$ unambiguously
from the data requires a measurement of the angular distribution,
which is difficult. 
In most experiments, it has been assumed that either $|G_{M}|= |G_{E}|$
(which should be a good approximation close to the two-nucleon threshold) 
or $|G_{E}|= 0$ (which should be a good approximation for large momentum
transfers). Most recent data have been presented
using the latter hypothesis.

The time-like data were already included in the dispersion analyses 
of Refs.~\cite{HMD96,slacproc}. The proton magnetic form factor up
to $t\approx 6$ GeV$^2$ was well described 
by these analyses. Data at higher momentum transfers were not included.
The data for the neutron magnetic form factor are from the pioneering
FENICE experiment \cite{Ant98}. They have been 
analyzed under both the assumption $|G_E|=|G_M|$ and 
$|G_E|=0$. The latter hypothesis is favored by the 
measured angular distributions \cite{Ant98}. Neither data set could be
described by the analysis \cite{slacproc}.

\begin{figure}[ht]
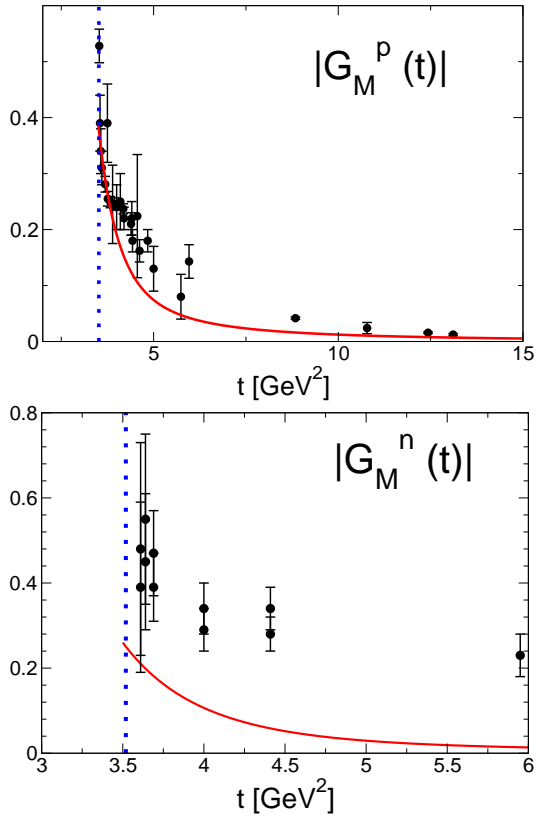
 
\centerline{
\includegraphics*[width=7.cm,angle=-0]{GMPt101205.eps}}
\centerline{
\includegraphics*[width=7.cm,angle=-0]{GMNt101205.eps}}
\caption{Current status of our analysis of the magnetic form factors
in the time-like region compared to the world data
\cite{protont1,protont2,protont3,protont4,protont5,Ant98}. 
The solid line gives our {\it preliminary best fit}, while the vertical dotted 
line indicates the two-nucleon threshold.
}
\label{timenew}
\end{figure}

In Fig.~\ref{timenew}, we show the current status of the analysis
of the time-like data for the magnetic form factors.
For the proton magnetic form factor, data up to momentum transfers
$t\approx 15$ GeV$^2$ have been included
\cite{protont1,protont2,protont3,protont4,protont5}.
Our preliminary fit gives a good description in 
the threshold region but starts to deviate significantly around
$t\approx 5$ GeV$^2$. The data for $t\geq 10$ GeV$^2$ are well 
described. This seems to 
due to a slight inconsistency in the data around 5 GeV$^2$ and for 
$t\geq 10$ GeV$^2$. This question deserves further attention.

The status for the neutron form factor is the same as in the previous 
analysis \cite{slacproc}: Neither of the two data sets 
from Ref.~\cite{Ant98} can be described. 
Even though we are not yet in the region where perturbative QCD 
is applicable, it comes as a surprise
that the neutron form factor is larger in magnitude than the 
proton one. Perturbative QCD predicts asymptotically equal magnitudes.
In any case, there is interesting physics in the time-like nucleon form 
factors and new precision experiments such as the PANDA and PAX
experiments at GSI would be very welcome.

\section{Pion Cloud of the Nucleon}
\label{sec:picloud}

Friedrich and Walcher (FW), recently analysed the em nucleon 
form factors and performed
various phenomenological fits \cite{FW}. Their fits showed a pronounced
bump structure in $G_E^n$ which they interpreted using an ansatz for the 
pion cloud based on the idea that the proton can be thought of as virtual 
neutron-positively charged pion pair. 
They found a very long-range contribution to the charge distribution 
in the Breit frame extending out to about 2~fm which they attributed to 
the pion cloud. 
While naively the pion Compton wave length is of this size, 
these findings are indeed surprising if compared with the ``pion cloud'' 
contribution due to the $2\pi$ continuum
contribution to the isovector spectral 
functions discussed in Sec.~\ref{sec:2pico}.

As was shown by Hammer, Drechsel, and Mei\ss ner \cite{Hammer:2003qv},
these latter contributions to  the long-range part of the nucleon 
structure are much more confined in coordinate space 
and agree well with earlier (but less systematic) calculations based on chiral 
soliton models, see e.g. \cite{UGM}. 
In the dispersion-theoretical framework, the longest-range part of the 
pion cloud contribution to the nucleon form factors is given
by the $2\pi$ continuum -- the lowest-mass intermediate 
state including only pions.
Note that a one-pion intermediate state is forbidden by parity.

The nonresonant part of the $2\pi$ continuum can be calculated in
ChPT \cite{Norbert} while the full continuum
can be obtained from experimental data and unitarity as discussed 
in  Sec~\ref{sec:2pico}. The \lq\lq pion cloud'' corresponds to the
nonresonant part of the $2\pi$ continuum excluding the $\rho$.
Consequently, the $\rho$ contribution has to be subtracted from the 
full $2\pi$ continuum.\footnote{Note that this separation is not unique.
It is only meaningful for the long-range part. The separation of the 
short-range part is model- and even representation-dependent.} 
The error in this subtraction was estimated using three different
methods for the separation of the contributions \cite{Hammer:2003qv}.

The charge distribution can be then be obtained
from the nonresonant part of the $2\pi$ continuum by Fourier
transformation. This leads to the relation:
\beq
\rho_i^v(r)=\frac{1}{4\pi^2}
\int_{4 M_\pi^2}^{40 M_\pi^2} dt
   \,{\rm Im}\,G_i^v(t) \,\frac{e^{-r\sqrt{t}}}{r}\,,
\eeq
where $i=E,M$. The contribution from $t\geq 40 M_\pi^2$ is small and can
be neglected \cite{Hammer:2003qv}.

\begin{figure}[ht] 
\includegraphics*[width=8.cm,angle=0]{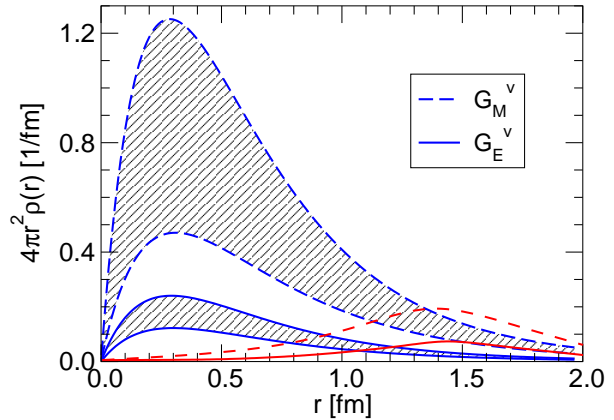}
\caption{Pion cloud contribution to the nucleon charge density.
The lines show the result of Friedrich and Walcher \cite{FW}, while
the bands give the result of Ref.~\cite{Hammer:2003qv}. Only the 
long-range contributions for $r\simgt 1$ are meaningful.
}
\label{rhoEMFW}
\end{figure}

The corresponding
result for the pion cloud contribution to the nucleon charge density
is shown in Fig.~\ref{rhoEMFW}.
The lines show the result of FW \cite{FW}, while
the bands give the result of Ref.~\cite{Hammer:2003qv}. Only the 
long-range contributions for $r\simgt 1$ fm are meaningful.
The separation of the short-range part into resonant
and nonresonant contributions is arbitrary. 
In comparison with Ref.~\cite{FW}, 
the $2\pi$ continuum contribution to the charge density 
is generally much smaller at distances beyond 1 fm,
e.g., by a factor of 3 for $\rho_E^v (r)$ at $r=1.5$ fm.
We emphasize that this result is obtained from independent
physical information that determines the $2\pi$ continuum 
(pion form factor and $\pi\pi\to N\bar{N}$ amplitudes,
cf.~Sec.~\ref{sec:2pico}) and not from form factor fits.

As a consequence, it remains to be shown
how the proposed long-range pion cloud can be reconciled with
what is known from dispersion relations and ChPT.
In order to clarify this issue, we have performed various fits in 
order to understand what structures in the spectral function are required to
reproduce the bump in $G_E^n$.
We find that the structure can only be reproduced if additional low-mass
strength in the spectral function below $t \simlt 1$ GeV$^2$ is allowed
beyond the $2\pi$, $K\bar{K}$, and $\rho\pi$ continua and the 
$\omega$ pole. In the fits
of Secs.~\ref{sec:fits} and \ref{sec:time} such strength was explicitly
excluded.

\begin{figure}[ht] 
\includegraphics*[width=8.cm,angle=0]{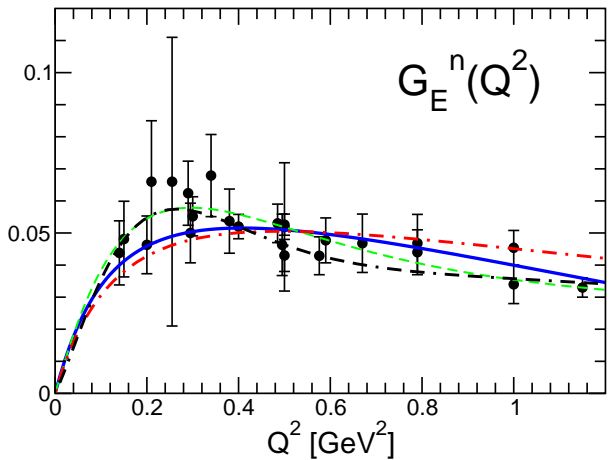}
\caption{The neutron electric form factor at low momentum transfer:
present fit with additional low-mass strength (dashed line) compared
to the fits of Friedrich and Walcher \cite{FW} (double-dash-dotted line).
For comparison, the fits of  Sec.~\ref{sec:fits} (solid line) and 
Ref.~\cite{HM04} (dash-dotted line) are also shown.
}
\label{Gendip}
\end{figure}

In Fig.~\ref{Gendip}, we show the neutron electric form factor at
low momentum transfer. The fit of FW \cite{FW} is given
by the double-dash-dotted line, while the present fit with 
additional low-mass strength is given by the dashed line.
For comparison, we show also the fit of  Ref.~\cite{HM04} (dash-dotted line)
and the fit from  Sec.~\ref{sec:fits} (solid line).
The fit with additional low-mass strength shows a clear bump structure
around $Q^2\sim 0.3$ GeV$^2$.
This structure requires three additional low mass poles:
two isoscalar poles at $M_s^2=0.13$~GeV$^2$, 0.54~GeV$^2$ and
one isovector pole at $M_v^2=0.30$~GeV$^2$.
In principle, vector meson dominance works well for $t\leq 1$ GeV$^2$
and one should be able to interpret these poles as physical vector
mesons. However, no such vector mesons are known in this region.
This raises the question of whether the effective low-mass poles can
be interpreted as something else?

One possible solution would be to interpret the poles as effective
poles mimicking some continuum contribution. It is interesting to 
note that the three low-mass poles happen to come out at the thresholds
of the $3\pi$, $4\pi$ and $5\pi$ continua and are located
in the correct isospin channel. 
Maybe these higher-order pion continua are more important than
previously thought and have a threshold enhancement similar to the
$2\pi$ continuum that is accounted for by the effective poles?

Even though this scenario has a certain appeal, it appears unlikely 
given the current state of knowledge.
In Ref.~\cite{BKMspec}, the threshold behavior of the  $3\pi$ continuum
was explicitly calculated in heavy baryon ChPT and no enhancement was found.
Moreover, the  inelasticity from four pions in $\pi\pi$ scattering
and four-pion production in $e^+ e^-$ annihilation at low momentum
transfer are known to be small \cite{LB,Gasser:1990bv,Ecker:2002cw}.

\section{Summary \& Outlook}

Dispersion theory simultaneously describes all four nucleon 
form factors over the whole range of momentum transfers in both the 
space-like and time-like regions. It allows for the inclusion of
constraints from other physical processes, unitarity, and ChPT
and therefore is an ideal tool to analyze the form factor data.

We have presented preliminary results for our new dispersion
analysis that is currently carried out in Bonn. The spectral function
has been improved and  contains the updated $2\pi$ continuum 
\cite{Belushkin:2005ds}, as well the $K\bar{K}$ 
\cite{Hammer:1998rz,Hammer:1999uf} and $\rho\pi$ 
continua \cite{Meissner:1997qt}. Our preliminary best
fit gives a consistent description of the world data in the 
space-like region. The understanding of the time-like form factors is more
difficult and a future challenge for theorists and experimentalists
alike. 

As part of this ongoing theoretical program, many things remain
to be done:

The stability constraint requires to use the minimum number of poles.
Our strategy for the future is to successively reduce the number of poles 
without sacrificing the quality of the fit.
Furthermore, the description of the time-like data needs to be improved.
In previous experiments, the separation of $G_E$ and $G_M$ could only be 
carried out under overly simplifying assumptions. New data, such as 
planned for the PANDA and PAX experiments at GSI, are therefore called for.

Other improvements concern the quantification of
theoretical and systematic uncertainties in the analysis,
the inclusion of perturbative QCD corrections beyond superconvergence
(leading logarithms etc.), and the inclusion of two-photon physics.
The latter point might require to analyze the cross section data directly.
Last but not least, the consequences of the new data for the 
strange vector form factors of the nucleon need to be worked out.

\section{Acknowledgments}

This work was done in collaboration with M.B. Belushkin,
D. Drechsel, and Ulf-G. Mei\ss ner. M.J. Ramsey-Musolf has contributed 
in earlier stages of the project. 

The work was supported in part by the EU I3HP under contract number 
RII3-CT-2004-506078 and the DFG through funds provided
to the SFB/\-TR 16 \lq\lq Subnuclear Structure of Matter''
and SFB 443 \lq\lq Many Body Structure of Strongly Interacting Systems''.

I would like to thank Hartmuth Arenh\"ovel, Hartmut Backe,
Dieter Drechsel, J\"org Friedrich, Karl-Heinz Kaiser, 
and Thomas Walcher for a very stimulating and enjoyable time in Mainz.
I have had many personal interactions with them through scientific
discussions and/or through lectures and seminars I attended as a student.
In particular, I want to thank my PhD advisor Dieter Drechsel from
whom I have learned much about physics and research.

\end{document}